\documentclass[aps,pre,floats,preprint,showpacs]{revtex4}
\usepackage{latexsym,amsmath}
\usepackage{amsbsy}
\usepackage[usenames]{color}
\usepackage{amssymb}
\usepackage{graphicx}

\begin{document}

\title{Rich-club vs rich-multipolarization phenomena in weighted networks}

\author{M. \'Angeles Serrano}
\affiliation{Instituto de F\'{i}sica Interdisciplinar y Sistemas Complejos IFISC (CSIC-UIB),\\
Campus Universitat Illes Balears, E-07122 Palma de Mallorca, Spain}

\date{\today}

\begin{abstract}
Large scale hierarchies characterize complex networks in different domains. Elements at their top, usually the most central or influential, may show multipolarization or tend to club forming tightly interconnected communities. The rich-club phenomenon quantified this tendency based on unweighted network representations.
Here, we define this metric for weighted networks and discuss the appropriate normalization which preserves nodes' strengths and discounts structural strength-strength correlations if present. We find that in some real networks the results given by the weighted rich-club coefficient can be in sharp contrast to the ones in the unweighted approach. We also discuss that the scanning of the weighted subgraphs formed by the high-strength hubs is able to
unveil features contrary to the average: the formation of
local alliances in rich-multipolarized environments, or a lack of
cohesion even in the presence of rich-club ordering. Beyond structure, this analysis matters for understanding correctly functionalities and dynamical processes relying on hub interconnectedness.
\end{abstract}


\maketitle

\section{Introduction}
A common feature of many real systems is a strongly
hierarchical organization which arises at the large scale as a
consequence of their microscopic dynamics. As a reflection, many structural
properties describing these systems are far from uniform and show an
extreme dispersion which marks a few of their elements with the
highest values as dominant. It is common to extrapolate their
prevalence beyond structure to recognize them as the most central, influential,
or primal in general terms. In what manner do these top elements relate to
each other, in particular whether they are polarized or, on the
contrary, show a tendency to club forming elites or backbones, is an
open question that matters for understanding the makeup and
performance of the whole system.

In the context of complex network science~\cite{Albert:2002}, the
quantitative discussion of this issue is known as the rich-club
phenomenon. In the network conceptualization of a real system, the
most fundamental statistic associated to the elements represented as
nodes is the number of neighbors they are connected to, the degree
$k$. In a vast majority of real networks the degree distributions
$P(k)$ is very broad and defines a topological hierarchy with
``rich-nodes'', those with a high degree, at the top. To detect if
they aggregate in a well interconnected core, a first uniparametric
measure, the rich-club coefficient, was proposed as the fraction of
edges actually connecting nodes with degree larger than a certain
threshold $k_T$ out of the maximum number of connections the nodes
in this subset would share in a perfect clique~\cite{Zhou:2004}.
Later on, the original metric was redesigned in order to discount
structural effects forcing hubs to be connected without the
intervention of special ordering principles~\cite{Colizza:2006b}. In
this way, rich-clubs have been found in scientific collaboration
networks and in critical infrastructures such as the world air
transportation system~\cite{Colizza:2006b}, or in the protein
interaction map of the human malaria parasite~\cite{Wuchty:2007}.

However, the first approximation of taking the interactions between
pairs of elements $i$ and $j$ as binary, just present or absent as described by the adjacency matrix $a_{ij}$, turns out to be
an oversimplification that in many analysis can distort the
interpretation of the results. An interaction can more exactly be quantified by its intensity or
weight, $w_{ij}$, and a node better be characterized by its strength, $s_i$, giving the actual intensity of the interactions it handles and defined as the sum of the weights on the links attached to it. Here, we explain how to generalize the concept of rich-club ordering to evaluate weighted networks~\cite{Newman:2001aa,Barrat:2004b}. More specifically, our work makes the following contributions:
\begin{itemize}
\item We define the rich-club coefficient in weighted networks as a function of the total weight of links connecting nodes of strength larger than a certain value as compared to the appropriate normalization. Section~\ref{sec2a}.
\item To compute the normalization, we introduce a new null model for weighted networks that preserves nodes' strengths but otherwise produces maximally random networks. This null model can provide a reference value in weighted networks beyond the rich-club coefficient. Subsection~\ref{sec2a}.
\item We compute analytically the strength based rich-club coefficient in the uncorrelated limit. Subsection~\ref{sec2b}.
\item We explore the role of structural strength-strength correlations by introducing two new metrics: the average nearest neighbors strength and the weighted average nearest neighbors strength. Subsection~\ref{sec2c}.
\item We apply the methodology to three different real networks and found that the consideration of weights can bring results in sharp contrast to those obtained from the degree-based measure. Subsection~\ref{sec3}.
\item We also discuss that the measurement of the rich-club coefficient is not enough for a complete assessment of the rich-club phenomenon. The averaging character of the metric hides the particulars about the internal organization of the subsets gathering the richest nodes. Their direct inspection, comparing the real weight in each inner link with the reference value given by the null model, uncovers features sometimes contrary to what the overall measure is implying: very fragmented subsets in the presence of rich-club ordering or local alliances when the dominant trend is rivalry or multipolarization (avoidance of connecting to each other in a group formed by several). Subsection~\ref{sec4}.
\end{itemize}
Conclusions can be found in Section~\ref{sec5}.

\section{Detecting rich-club ordering in weighted networks}\label{sec2}
We begin by noticing that the degree-based rich-club coefficient $\rho(k_T)$ of a
given graph is computed in two recursive steps. In the first, a
simple degree thresholding procedure is applied to produce a
hierarchy of nested subgraphs formed of nodes with degrees larger
than an increasing threshold $k_T$~\footnote{This nested hierarchy of subgraphs turns out
to have self-similarity properties for some real scale-free
networks such as the Internet at the autonomous system level, see~\cite{Serrano:2008a}.}. In the second, the number of
connections within each subgraph $k>k_T$ is evaluated and
compared against the corresponding value in the randomized version
of the graph that preserves the degree distribution
$P(k)$~\cite{Maslov:2002}. Hence, the rich-club
coefficient can be written as~\cite{Colizza:2006b}
\begin{equation}
\rho(k_T)=\frac{E_{k>k_T}}{E_{k>k_T}^{ran}}.
\end{equation}
A ratio larger than one indicates the presence of a
rich-club phenomenon, high-degree nodes being
intertwined with one another more tightly than expected from randomness. In contrast, a ratio lesser than one is a signature of an
opposite organizing principle that leads to a lack of
interconnectivity among high-degree nodes.

\subsection{Null model which preserves the strength distribution and definition of the rich-club coefficient}\label{sec2a}
The computational procedure for weighted networks
needs to redefine the subgraphs and the appropriate null model. There is not a unique choice, since one could for instance be interested in a quenched topology where only weights are randomized keeping nodes' strengths constant~\cite{Bhattacharya:2008}. Here, we are however interested in avoiding the constraint of the actual topology of the real network in order to detect departures from the random counterpart not only in terms of intensities but even regarding the presence or absence of interactions. A link between hubs predicted by the null model that is actually missing in the real network is clearly indicating a tendency contrary to club.

Hence, we focus exclusively on strengths and assume rich nodes as those with the highest values. The appropriate thresholding procedure applied to the weighted networks generates then a hierarchy of nested subgraphs
of nodes with strengths larger than an increasing threshold $s_T$, and the sum of the weights
on the links within each subgraph, $W_{s>s_T}$, is considered. Regarding the normalization, we propose to compare to the randomized version of the graph which preserves the strength distribution $P(s)$. This
null model can be achieved by approximating the weights in the
network by integers, so that they could be considered as multiple
connections formed by decoupled links. Then, the usual randomization
based on rewirings~\cite{Maslov:2002} can be done avoiding self-connections
but not multiple ones. In this way, the nodes maintain their
strength but the weights in the links (or the degrees) can change in
the process. To avoid inducing correlations, notice that each decoupled link should be selected independently with the same probability. This ensures that in the steady state the weights agree with the expected values in a strength preserving but otherwise maximally random conformation. Formally, the rich-club coefficient in the weighted approach can be written as
\begin{equation}
\rho(s_T)=\frac{W_{s>s_T}}{W_{s>s_T}^{ran}}.
\label{rhow}
\end{equation}

\subsection{Analytical computation of the uncorrelated limit in the weighted approach}\label{sec2b}
Strengths are less prone than degrees to be
affected by structural
constraints~\cite{Boguna:2004,Serrano:2005c}. This fact makes
meaningful the consideration of the uncorrelated limit
in the weighted approach,
\begin{equation}
\rho^{unc}(s_T)=W_{s>s_T}/W_{s>s_T}^{unc},
\label{rhounc}
\end{equation}
in which the strengths of attached nodes are independent. The normalization $W_{s>s_T}^{unc}$ can be computed
analytically. In the same spirit of the original measure of Zhou and
Mondragon~\cite{Zhou:2004}, it is given by the sum of the
uncorrelated weights in the fully connected subsets, which can be
calculated just by taking into account that, in average, they must
be proportional to the product of the strengths of the nodes $i$ and
$j$ they are associated to~\cite{Serrano:2005c}. If loops are not
allowed,
\begin{equation}
W_{s>s_T}^{unc}=\sum_{\nu_i}\sum_{\nu_{j \neq i}} w_{ij}^{unc}=\left< s \right>\frac{\sum_{\nu_i}\sum_{\nu_j \neq i}s_i s_j}{N
\left< s \right>^2 - \left< s^2 \right>},
\end{equation}
where $\nu_i$ designates the subsets of nodes such that $s_i> s_T$, $N$ is the total number of nodes in the network, and $\left< s \right>$ and $\left< s^2 \right>$ are the first and second moments of the strength distribution.

\subsection{Structural strength-strength correlations}\label{sec2c}
As it happens for the degrees~\cite{Boguna:2004}, in some networks closure conditions enforce the presence of correlations between the strengths of connected pairs that cannot be avoided even in maximally random configurations. Equation~\ref{rhow} discounts these structural strength–-strength correlations
and other higher-order effects which are not polished off by the random
procedure. However, the uncorrelated approximation assume a total absence of dependencies between strengths and it is not perfectly valid in structural strength-strength correlations are present. To help to discern in which networks these are important, we define --from the formalism in~\cite{Serrano:2006d} and in analogy to the average nearest neighbors degree~\cite{Pastor-Satorras:2001}-- the average nearest neighbors strength and the weighted average nearest neighbors strength, both as a function of the strength,
\begin{equation}
\bar{s}_{nn}(s)=\frac{1}{N_s}\sum_{i\in s, j} \frac{a_{ij}}{k_i}s_j, \hspace{0.25cm}
\bar{s}^w_{nn}(s)=\frac{1}{N_s}\sum_{i\in s, j} \frac{w_{ij}}{s}s_j,
\end{equation}
being $N_s$ the number of nodes with strength $s$. The average $\bar{s}_{nn}(s)$ is coupled to the underlying degree structure and computes strength-strength correlations that are structural and cannot be destroyed by the randomization. In contrast, $\bar{s}^w_{nn}(s)$ is disentangled from degrees and expected to be perfectly flat in the maximally random case. Both measures combined enable the discrimination of structural and non-structural strength-strength correlations in weighted networks, setting-up the validity of the uncorrelated approximation.

\section{Real networks}\label{sec3}
\begin{figure*}[t]
\includegraphics[width=17.8cm]{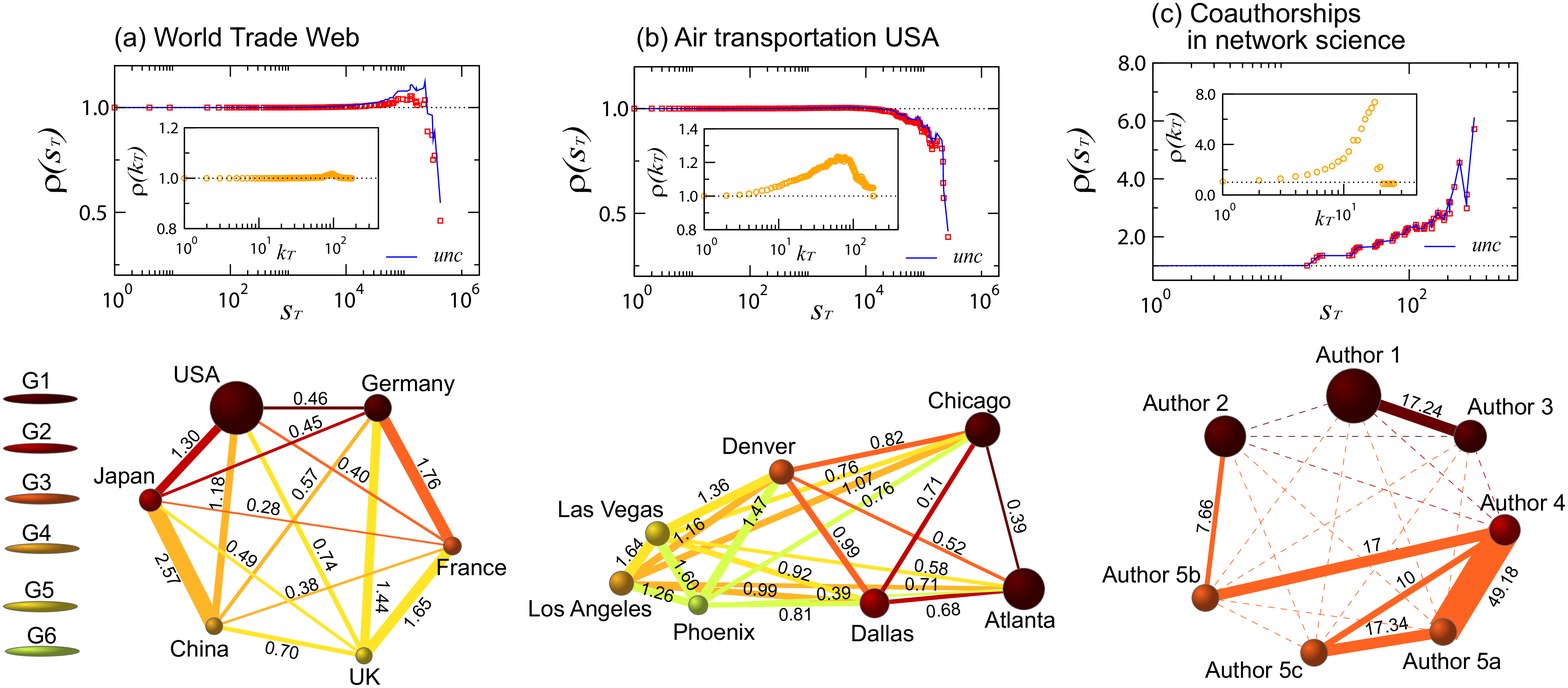}
\caption{Weighted rich--club phenomenon in the real networks WTW, USAN and CCNN. AT the top, graphs for the strength-based rich-club coefficient. Dotted curves correspond to $\rho(s_T)$ normalized averaging over 100 randomizations (rewirings) of the original network. Weights have been discretized in all cases: rounded off to the nearest integer in the WTW, coarse-grained to hundreds and rounded off to the nearest
integer in the USAN, and divided by the minimum weight and rounded off to the nearest integer in the CNCN. Solid lines in the plots represent the analytical curves for the uncorrelated approximation $\rho^{unc}(s_T)$. The insets show the degree-based rich-club $\rho(k_T)$ of the unweighted representation. At the bottom, sketches showing greater detail of how the hubs interact between them. Darker colors represent inner subsets in the nested hierarchy as defined by the threshold strength. Within a plot, the sizes of the nodes are proportional to their strengths. The numerical values labeling the links represent the ratio of the actual weight of the tie to its average value in the randomized versions.}
\label{fig1}
\end{figure*}
In the graphs of Fig.~1 (top), we report the behavior of $\rho(s_T)$ and $\rho^{unc}(s_T)$ in three different real networks, and for comparison we also provide
the curves for the degree-based definition $\rho(k_T)$ (insets).
These examples correspond to: (a) the world trade web (WTW) of commercial
relationships between states~\cite{Serrano:2003} in 2000, where the
weights give the annual merchandize exchanges in millions of
current-year US dollars (http://weber.ucsd.edu/~kgledits/exptradegdp.html~\cite{Gleditsch:2002})
(b) the domestic segment of the USA
airport network (USAN) for
the year 2006 (http://www.transtats.bts.gov/), where the weights are
given by number of passengers~\cite{Barrat:2004b}, and (c) an extract of the actual network of coauthorships between
researchers in the area of Complex Networks (CCNN)~\cite{Newman:2006a},
where the weights represent the intensity of the collaborative ties
depending on the number coauthored papers and number of authors in each. To validate our methodology, we also generated a maximally random network at the weighted level with a size of $2 \times 10^4$ nodes and a strength distribution $P(s)\sim s^{-1.85}$ making use of the weighted configuration model (WCM), see~\cite{Serrano:2005c} and references therein. By construction and as expected, its strength-based rich-club coefficient does not detect any ordering but has a value of one in the whole domain (plot omitted for brevity).

Surprisingly, very different behaviors of $\rho(k_T)$ and $\rho(s_T)$ can be detected in networks where de degree and the strength are not trivially related. Two instances are the WTW and the USAN, which respectively have a neutral --meaning that the unweighted representation is dominated by structural connectivity effects-- and a mild degree-based rich-club ordering but exhibit a decreasing strength-based coefficient providing evidence of a clear rich-multipolarization phenomenon. Their oligarchies of rich nodes are on average loosely interconnected in terms of weight as compared with the random null model counterpart, in contraposition to the rich-club situation. On the other hand, the degree-based and strength-based spectrums of homogeneous networks --those with weights uncorrelated with degrees-- are expected to be qualitatively similar. This is what happens in the CCNN case. The presence of a strong rich-club ordering in both the unweighted and the weighted representations seems to provide support to the idea that the more collaborative --and extrapolating maybe the more influential-- researchers in complex network science tend to club following a expected tendency in social systems (as we will discuss below, maybe excluding the very top hubs as suggested by the sharp decay of $\rho(k_T)$ for very high degrees).

Turning now to evaluate the goodness of fit between the uncorrelated approximation $\rho^{unc}(s_T)$ and $\rho(s_T)$, they match almost perfectly in the USAN and the CCNN. The agreement is slightly worse for the WTW and in the case of the simulated WCM the approximation is clearly bad (plot omitted for brevity). The explanation to this divergence can be found in the presence of structural strength-strength correlations. In Fig.~2, we report these functions for the real and simulated networks. Disassortative~\cite{Newman:2002a} structural strength correlations measured by $\bar{s}_{nn}(s)$ are important for the WTW but much more the WCM, limiting the validity of the uncorrelated approximation in these cases.
\begin{figure}[t]
\includegraphics[width=8.6cm]{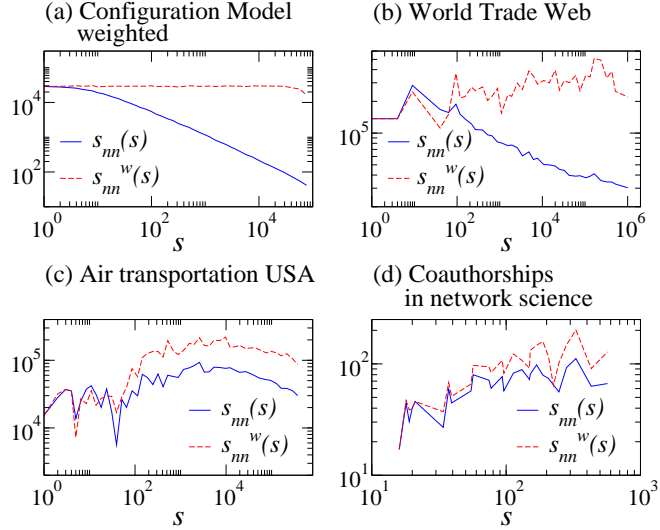}
\caption{Weighted strength-strength correlation structure of the real and the simulated networks. $\bar{s}_{nn}(s)$ for the WTW and the WCM are power-laws with negative exponents, larger in the later case, which denotes disassortative strength mixing.} \label{fig3}
\end{figure}

\section{Analysis of the inner subgraphs}\label{sec4}
We have now a methodology to detect rich-club ordering in weighted networks which provides valuable information about the average tendency in the interactions between hubs. But averages can hide local regularities in contrast to the overall behavior and a thorough assessment of the property requires greater detail. A direct inspection of the subgraphs formed by the hubs comparing the actual weight in each link with the reference value given by the null model is necessary. If, for a given link, the ratio of the two values is above one the two nodes connected by that link show a tendency to interact, while a ratio below one denotes the opposite tendency to avoid interaction. Besides, the stronger polarization is manifested when a link between two nodes is absent in the real network but it has a certain expected weight according to the null model.

The results for the real networks under study are summarized in the sketches of Fig.~1 (bottom). The WTW shows for instance a very clear and sharp rich-multipolarization phenomenon. Function $\rho(s_T)$ runs at value one just until the tail end, where a group of rich nodes --UK, China, France, Japan, Germany, and USA, in increasing order of strength-- form subsets with $\rho(s_T)<1$. Notice the overlap of five members --all except China-- with the club of the seven largest industrialized and richest countries in the world, the G7, which also includes Italy and Canada (the next two countries in the strength hierarchy of the WTW in 2000). The decreasing tail indicates that these hub countries share on average less weight among them that the expected in the random situation. So they seem to have an aversion to connect to each other as anticipated between powers in a strongly competitive economic system. However, they interact anyway forced by structural constraints and form fully connected subsets. What is more interesting, the biggest world economies seem to be polarized into two connected blocks in direct competition: the USA and its Asian allies Japan and China, against Europe, France and Germany including UK. Each block is tightly connected by trading volumes larger than random predictions and competition between the two is exposed by a reduced exchange of merchandizes as compared to the null model.

The situation for the USAN is similar in the sense that it presents a distinct rich-multipolarization phenomenon (although the transition from $\rho(s_T)=1$ to $\rho(s_T)<1$ is smoother) and that the hub airports form fully connected subgraphs. In the sketch we show the seven largest airports in number of domestic passengers. Again, although the overall tendency is a reduced interaction reflecting competition, the examination of the subsets discovers tight interconnection among the big west coast and central airports, Los Angeles, Phoenix, Las Vegas and Denver, and a neutral interaction with Dallas, while the interactions with Chicago and Atlanta in the east coast are weak. Chicago and Atlanta are the biggest hubs with a weak tie to each other and to the rest. An explanation can be found in the geographical layout. While Chicago and Atlanta are far away and surrounded by not overlapping basins of attraction of smaller airports and passengers, Los Angeles, Phoenix, and Las Vegas are close enough to be forced to share approximatively the area of influence so that they interact to each other more than expectable. Surely, other aspects such as the provision of connections to international flights or operational constraints of the biggest air companies can also have a role.

Finally, the CCNN illustrates the opposite situation. This is a network with a strong rich-club ordering both in the degree-based and the strength-based approach which seems to suggest that researchers tend to form collaborative groups. However, this does not necessarily mean a high degree of cohesiveness among all the most collaborative scientists. Indeed, the inner subgraphs of the network are actually very sparse even though they are expected to be fully connected according to the random null model. This points out to a certain level of rivalry between the hubs. They form ties with just very few other hubs, but in exchange the interactions are very strong and enough to decide the averages.

We would like to emphasize that, in all three cases, these results are corroborated and in good agreement with other sources of information about the system: commercial agreements in the case of the WTW, geographical layout in the case of the airports network, and strong competitive character of collaborations in the field of complex networks science.

\section{Conclusions}\label{sec5}
These findings have important consequences. At the theoretical level, detecting rich-club ordering is dependent on whether the intensities of the interactions among elements are taken into account in systems where intensities and number of interactions are related in a non-trivial form. Besides, an exhaustive assessment of the property requires greater detail that the one that can be achieved by the averaging coefficient. The scanning of the subgraphs formed by the hubs as compared to the appropriate null model is also relevant and uncovered the formation of local alliances in multipolarization environments or a lack of cohesion even in the presence of rich-club ordering. Beyond structure, this analysis
matters for understanding functionalities and dynamical processes relying on hub interconnectedness
and, in a broader context, may help explain how primary forces such as competition and
cooperation influence collective form and performance.

\begin{acknowledgments}
This work was supported by DELIS FET Open 001907, the SER-Bern
02.0234, and by DGES grant FIS2007-66485-C02-01. The author thanks
Mari\'{a}n Bogu\~{n}\'{a}, Paolo De Los Rios, and A. Vespignani for useful discussions.
\end{acknowledgments}

{\it Note added.}—- Shortly after this paper was finished, another group proposed an alternative null model in order to detect rich-club ordering in weighted complex networks, see~\cite{Opsahl:2008}.


\begin{thebibliography}{17}
\expandafter\ifx\csname natexlab\endcsname\relax\def\natexlab#1{#1}\fi
\expandafter\ifx\csname bibnamefont\endcsname\relax
  \def\bibnamefont#1{#1}\fi
\expandafter\ifx\csname bibfnamefont\endcsname\relax
  \def\bibfnamefont#1{#1}\fi
\expandafter\ifx\csname citenamefont\endcsname\relax
  \def\citenamefont#1{#1}\fi
\expandafter\ifx\csname url\endcsname\relax
  \def\url#1{\texttt{#1}}\fi
\expandafter\ifx\csname urlprefix\endcsname\relax\def\urlprefix{URL }\fi
\providecommand{\bibinfo}[2]{#2}
\providecommand{\eprint}[2][]{\url{#2}}

\bibitem[{\citenamefont{Albert and Barab{\'a}si}(2002)}]{Albert:2002}
\bibinfo{author}{\bibfnamefont{R.}~\bibnamefont{Albert}} \bibnamefont{and}
  \bibinfo{author}{\bibfnamefont{A.-L.} \bibnamefont{Barab{\'a}si}},
  \bibinfo{journal}{Rev. Mod. Phys.} \textbf{\bibinfo{volume}{74}},
  \bibinfo{pages}{47} (\bibinfo{year}{2002}).

\bibitem[{\citenamefont{Zhou and Mondragon}(2004)}]{Zhou:2004}
\bibinfo{author}{\bibfnamefont{S.}~\bibnamefont{Zhou}} \bibnamefont{and}
  \bibinfo{author}{\bibfnamefont{R.~J.} \bibnamefont{Mondragon}},
  \bibinfo{journal}{IEEE Commun. Lett.} \textbf{\bibinfo{volume}{8}},
  \bibinfo{pages}{180} (\bibinfo{year}{2004}).

\bibitem[{\citenamefont{Colizza et~al.}(2006)\citenamefont{Colizza, Flammini,
  Serrano, and Vespignani}}]{Colizza:2006b}
\bibinfo{author}{\bibfnamefont{V.}~\bibnamefont{Colizza}},
  \bibinfo{author}{\bibfnamefont{A.}~\bibnamefont{Flammini}},
  \bibinfo{author}{\bibfnamefont{M.~A.} \bibnamefont{Serrano}},
  \bibnamefont{and}
  \bibinfo{author}{\bibfnamefont{A.}~\bibnamefont{Vespignani}},
  \bibinfo{journal}{Nature Physics} \textbf{\bibinfo{volume}{2}},
  \bibinfo{pages}{110} (\bibinfo{year}{2006}).

\bibitem[{\citenamefont{Wuchty}(2007)}]{Wuchty:2007}
\bibinfo{author}{\bibfnamefont{S.}~\bibnamefont{Wuchty}},
  \bibinfo{journal}{PLoS ONE} \textbf{\bibinfo{volume}{2}},
  \bibinfo{pages}{e335} (\bibinfo{year}{2007}).

\bibitem[{\citenamefont{Newman}(2001)}]{Newman:2001aa}
\bibinfo{author}{\bibfnamefont{M.~E.~J.} \bibnamefont{Newman}},
  \bibinfo{journal}{Phys. Rev. E} \textbf{\bibinfo{volume}{64}},
  \bibinfo{pages}{016132} (\bibinfo{year}{2001}).

\bibitem[{\citenamefont{Barrat et~al.}(2004)\citenamefont{Barrat,
  Barth\'{e}lemy, Pastor-Satorras, and Vespignani}}]{Barrat:2004b}
\bibinfo{author}{\bibfnamefont{A.}~\bibnamefont{Barrat}},
  \bibinfo{author}{\bibfnamefont{M.}~\bibnamefont{Barth\'{e}lemy}},
  \bibinfo{author}{\bibfnamefont{R.}~\bibnamefont{Pastor-Satorras}},
  \bibnamefont{and}
  \bibinfo{author}{\bibfnamefont{A.}~\bibnamefont{Vespignani}},
  \bibinfo{journal}{Proc. Natl. Acad. Sci. USA} \textbf{\bibinfo{volume}{101}},
  \bibinfo{pages}{3747} (\bibinfo{year}{2004}).

\bibitem[{\citenamefont{Maslov and Sneppen}(2002)}]{Maslov:2002}
\bibinfo{author}{\bibfnamefont{S.}~\bibnamefont{Maslov}} \bibnamefont{and}
  \bibinfo{author}{\bibfnamefont{K.}~\bibnamefont{Sneppen}},
  \bibinfo{journal}{Science} \textbf{\bibinfo{volume}{296}},
  \bibinfo{pages}{910–913} (\bibinfo{year}{2002}).

\bibitem[{\citenamefont{Bhattacharya et~al.}(2008)\citenamefont{Bhattacharya,
  Mukherjee, Saram{\"a}ki, Kaski, and Manna}}]{Bhattacharya:2008}
\bibinfo{author}{\bibfnamefont{K.}~\bibnamefont{Bhattacharya}},
  \bibinfo{author}{\bibfnamefont{G.}~\bibnamefont{Mukherjee}},
  \bibinfo{author}{\bibfnamefont{J.}~\bibnamefont{Saram{\"a}ki}},
  \bibinfo{author}{\bibfnamefont{K.}~\bibnamefont{Kaski}}, \bibnamefont{and}
  \bibinfo{author}{\bibfnamefont{S.~S.} \bibnamefont{Manna}},
  \bibinfo{journal}{J. Stat. Mech.} \textbf{\bibinfo{volume}{P02002}}
  (\bibinfo{year}{2008}).

\bibitem[{\citenamefont{Bogu{\~n}\'{a}
  et~al.}(2004)\citenamefont{Bogu{\~n}\'{a}, Pastor-Satorras, and
  Vespignani}}]{Boguna:2004}
\bibinfo{author}{\bibfnamefont{M.}~\bibnamefont{Bogu{\~n}\'{a}}},
  \bibinfo{author}{\bibfnamefont{R.}~\bibnamefont{Pastor-Satorras}},
  \bibnamefont{and}
  \bibinfo{author}{\bibfnamefont{A.}~\bibnamefont{Vespignani}},
  \bibinfo{journal}{European Physical Journal B} \textbf{\bibinfo{volume}{38}},
  \bibinfo{pages}{205} (\bibinfo{year}{2004}).

\bibitem[{\citenamefont{Serrano and Bogu{\~n}\'{a}}(2005)}]{Serrano:2005c}
\bibinfo{author}{\bibfnamefont{M.~A.} \bibnamefont{Serrano}} \bibnamefont{and}
  \bibinfo{author}{\bibfnamefont{M.}~\bibnamefont{Bogu{\~n}\'{a}}},
  \bibinfo{journal}{AIP Conference Proocedings} \textbf{\bibinfo{volume}{776}},
  \bibinfo{pages}{101} (\bibinfo{year}{2005}).

\bibitem[{\citenamefont{Serrano and Bogu{\~n}\'{a}}(2003)}]{Serrano:2003}
\bibinfo{author}{\bibfnamefont{M.~A.} \bibnamefont{Serrano}} \bibnamefont{and}
  \bibinfo{author}{\bibfnamefont{M.}~\bibnamefont{Bogu{\~n}\'{a}}},
  \bibinfo{journal}{Phys. Rev. E} \textbf{\bibinfo{volume}{68}},
  \bibinfo{pages}{015101(R)} (\bibinfo{year}{2003}).

\bibitem[{\citenamefont{Gleditsch}(2002)}]{Gleditsch:2002}
\bibinfo{author}{\bibfnamefont{K.~S.} \bibnamefont{Gleditsch}},
  \bibinfo{journal}{J. Conflict Resolut.} \textbf{\bibinfo{volume}{46}},
  \bibinfo{pages}{712–724} (\bibinfo{year}{2002}).

\bibitem[{\citenamefont{Newman}(2006)}]{Newman:2006a}
\bibinfo{author}{\bibfnamefont{M.~E.~J.} \bibnamefont{Newman}},
  \bibinfo{journal}{Phys. Rev. E} \textbf{\bibinfo{volume}{74}},
  \bibinfo{pages}{036104} (\bibinfo{year}{2006}).

\bibitem[{\citenamefont{Serrano et~al.}(2006)\citenamefont{Serrano,
  Bogu{\~n}\'{a}, and Pastor-Satorras}}]{Serrano:2006d}
\bibinfo{author}{\bibfnamefont{M.~A.} \bibnamefont{Serrano}},
  \bibinfo{author}{\bibfnamefont{M.}~\bibnamefont{Bogu{\~n}\'{a}}},
  \bibnamefont{and}
  \bibinfo{author}{\bibfnamefont{R.}~\bibnamefont{Pastor-Satorras}},
  \bibinfo{journal}{Phys. Rev. E} \textbf{\bibinfo{volume}{74}},
  \bibinfo{pages}{055101(R)} (\bibinfo{year}{2006}).

\bibitem[{\citenamefont{Pastor-Satorras
  et~al.}(2001)\citenamefont{Pastor-Satorras, V\'{a}zquez, and
  Vespignani}}]{Pastor-Satorras:2001}
\bibinfo{author}{\bibfnamefont{R.}~\bibnamefont{Pastor-Satorras}},
  \bibinfo{author}{\bibfnamefont{A.}~\bibnamefont{V\'{a}zquez}},
  \bibnamefont{and}
  \bibinfo{author}{\bibfnamefont{A.}~\bibnamefont{Vespignani}},
  \bibinfo{journal}{Phys. Rev. Lett.} \textbf{\bibinfo{volume}{87}},
  \bibinfo{pages}{258701} (\bibinfo{year}{2001}).

\bibitem[{\citenamefont{Newman}(2002)}]{Newman:2002a}
\bibinfo{author}{\bibfnamefont{M.~E.~J.} \bibnamefont{Newman}},
  \bibinfo{journal}{Phys. Rev. Lett.} \textbf{\bibinfo{volume}{89}},
  \bibinfo{pages}{208701} (\bibinfo{year}{2002}).

\bibitem[{\citenamefont{Serrano et~al.}(2008)\citenamefont{Serrano, Krioukov,
  and Bogu{\~n}\'{a}}}]{Serrano:2008a}
\bibinfo{author}{\bibfnamefont{M.~A.} \bibnamefont{Serrano}},
  \bibinfo{author}{\bibfnamefont{D.}~\bibnamefont{Krioukov}}, \bibnamefont{and}
  \bibinfo{author}{\bibfnamefont{M.}~\bibnamefont{Bogu{\~n}\'{a}}},
  \bibinfo{journal}{Phys. Rev. Lett.} \textbf{\bibinfo{volume}{100}},
  \bibinfo{pages}{078701} (\bibinfo{year}{2008}).

\bibitem[{\citenamefont{Opsahl et~al.}(2008)\citenamefont{Opsahl, Colizza, Panzarasa and Ramasco}}]{Opsahl:2008}
\bibinfo{author}{\bibfnamefont{T.} \bibnamefont{Opsahl}},
  \bibinfo{author}{\bibfnamefont{V.}~\bibnamefont{Colizza}},
  \bibinfo{author}{\bibfnamefont{P.} \bibnamefont{Panzarasa}}, \bibnamefont{and}
  \bibinfo{author}{\bibfnamefont{J.~J.}~\bibnamefont{Ramasco}},
  \bibinfo{journal}{arXiv:0804.0417}.

\end{thebibliography}

\end{document}